# Dalton Transactions

## ARTICLE

## Defects in Metal-Organic Frameworks: A compromise between adsorption and stability?

A. W. Thornton,[a] R. Babarao,[a] A. Jain,[a,b] F. Trousselet[c] and F.-X. Coudert[c]

Defect engineering has arisen as a promising approach to tune and optimise the adsorptive performance of metal-organic frameworks. However, the balance between enhanced adsorption and structural stability remains an open question. Here both $CO_2$ adsorption capacity and mechanical stability are calculated for the zirconium-based UiO-66, which is subject to systematic variations of defect scenarios. Modulator-dependence, defect concentration and heterogeneity are explored in isolation. Mechanical stability is shown to be comprised at high pressures where uptake is enhanced with an increase in defect concentration. Nonetheless this reduction in stability is minimised for *reo* type defects and defects with trifluoroacetate substitution. Finally, heterogeneity and auxeticity may also play a role in overcoming the compromise between adsorption and stability.

## Introduction

Metal-Organic Frameworks (MOFs) or porous coordination polymers[1,2] have emerged with promising properties for industrial applications such as chemical separation, storage, sensing and catalysis but their mechanical, thermal, chemical and hydro stability have remained poor.[3]

Defect engineering[4] has arisen as an approach to further tune sorption, catalysis,[5-7] band gap,[8] magnetic[9] and electrical/conductive properties. Wu *et al.* showed that defects enhance pore volume, surface area and $CO_2$ uptake.[10] Ghosh *et al.* showed that $CO_2$ and $H_2O$ uptake could be enhanced with the controlled location and concentration of defects.[11] Shearer *et al.*[12] and other studies have shown that the amount and type of defects can be controlled with the choice of modulators and synthesis conditions.[7,13]

Defects are conventionally thought to reduce stability, however a few studies have shown that this depends on the type and nature of the defects. The stability of zirconium-based UiO-66 (and other UiO-series such as hafnium-based frameworks) has been attributed to the large number of coordination sites ($z$) per cluster, consisting of $z$ = 12 compared to $z$ = 6 for MOF-5 and $z$ = 4 for ZIF-8 and HKUST-1.[14] Defects in the form of missing linkers or missing clusters would likely reduce the overall stability, however De Voorde *et al.* demonstrated that the replacement of a linker with a functional group (trifluoroacetate) improved the mechanical stability under ball milling treatment.[15] Other studies have shown that defects can propagate to form defect planes in 2D across the whole dimension of the crystal as seen with fluorescence microscopy.[16] Cliffe *et al.* further explored this correlation of defects and revealed the formation of nano-regions.[17] Though the effect of heterogeneity on sorption and stability remains an open question.

No attention has yet been made on the balance between enhancement in adsorption versus the effect on stability with the introduction of defects. Sholl and Lively raised the question whether defects were a challenge or an opportunity.[18] In the same spirit, the work presented here aims to understand the cost-benefit of defects by computationally exploring a systematic variation of defect concentration and type, introduced within UiO-66, see Figure 1. Although the scope of possible defects is infinite, this controlled case study provides the opportunity to target and expose the isolated effects of modulator-dependence, defect concentration and heterogeneity, that are translatable to many other systems. Finally, carbon dioxide is chosen as the adsorbate of interest because of the global importance in reducing greenhouse gas emissions from fossil fuel-based power generators.[19]

## Experimental

### Materials

The fully activated structural model of UiO-66 in its hydroxylated form was constructed from the X-ray diffraction data reported by Cavka *et al.*[20] UiO-66 consists of vertices (or nodes) represented by $Zr_6O_4(OH)_4$ clusters connected via linkers made of $C_6H_4(COO)_2$ ions or benzene-di-carboxylate (BDC). The $Zr_6$ clusters are 12-coordinated, and the framework is of the face-centered cubic type (*fcu*), with $F\bar{4}3m$ symmetry. The

[a.] *Manufacturing Flagship, Commonwealth Scientific and Industrial Research Organisation, Private Bag 10, Clayton Sth, VIC 3169, Australia. Email: Ravichandar.Babarao@csiro.au and Aaron.Thornton@csiro.au*
[b.] *Indian Institute of Technology, Kanpur Nagar, Uttar Pradesh, 208016, India.*
[c.] *PSL Research University, Chimie ParisTech – CNRS, Institut de Recherche de Chimie Paris, 75005 Paris, France.*

crystallographic cell of this structure is thus cubic, with cell parameters $a \sim 21$ Å. Since H atoms cannot be located from X-ray diffraction data, these atoms were added to the organic group and μ$_3$-O position to neutralize the overall structure. For constructing defective UiO-66, the number of missing linkers per cluster was varied and compensated with the different capping ligands including formate, acetate, chloride, trifluoroacetate and hydroxide ions. Taking the example of formate (HCOO) capped ligands (or modulators), each defect consists of the replacement of a BDC ligand by two formate ions, at both endpoints of the linkage. For this modulator, five types of defective structures were considered at the unit cell scale (e.g. 4 clusters and 24 ligands for non-defective UiO-66). Defect positions were chosen so that the cluster coordinence is uniform and all missing linkers are if possible collinear, or at least coplanar:

- 1 defect at each cluster (2 per nominal cell), i.e. the cluster coordinence is $z = 11$ and the corresponding structure has P$\bar{4}$3$m$ symmetry.
- 2 defects at each cluster (such that all missing linkers in the cell are collinear), i.e. the cluster coordinence is $z = 10$ and the structure has I$mm$2 symmetry.
- 3 defects at each cluster (all missing ligands are coplanar), i.e. the cluster coordinence is $z = 9$ and the structure has C$m$ symmetry.
- 4 defects at each cluster (all ligands in a given plane, say $xy$, are substituted), i.e. the cluster coordinence is $z = 8$ and the structure has I$mm$2 symmetry.
- 4 defects per cluster with a missing cluster (compared to the nominal structure). This corresponds to a framework called *reo*, and has P$\bar{4}$3$m$ symmetry. X-ray diffraction showed indications that this structure is more favourable than other structures with the same amount of defects.[17]

In addition, combinations of defects were considered, at the supercell scale (i.e. 64 clusters for non-defective UiO-66), for adsorption simulations only. Here six scenarios are generated where each supercell has an average of 9 linkers per cluster:

- Every cluster has exactly 9 linkers.
- Half of the clusters have 8 linkers and the other half have 10 linkers.
- Four scenarios where each supercell has 4 missing clusters (*reo* type defects) but with different degrees of adjacency from 1 to 4. Adjacency is defined here as the number of missing clusters directly adjacent from each other. For example, an adjacency of 2 means that there are 2 missing cluster sites directly next to each other.

Prior to performing adsorption simulations, geometric optimization were performed for the conventional unit cells of both the perfect and defective UiO-66 structures using Vienna ab initio simulation package (VASP)[21-23] with a plane-wave energy cut-off of 500 eV and a Gamma-point mesh for sampling the Brillouin zone. The density derived electrostatic and chemical (DDEC) method[24] was used to calculate the atomic charges based on a periodic unit cell for both perfect and defective UiO-66 structures using VASP software. However for the supercell scenarios, geometric optimization was performed using FORCITE module[25] based on the Universal Force Field (UFF)[26] due the large computational cost.

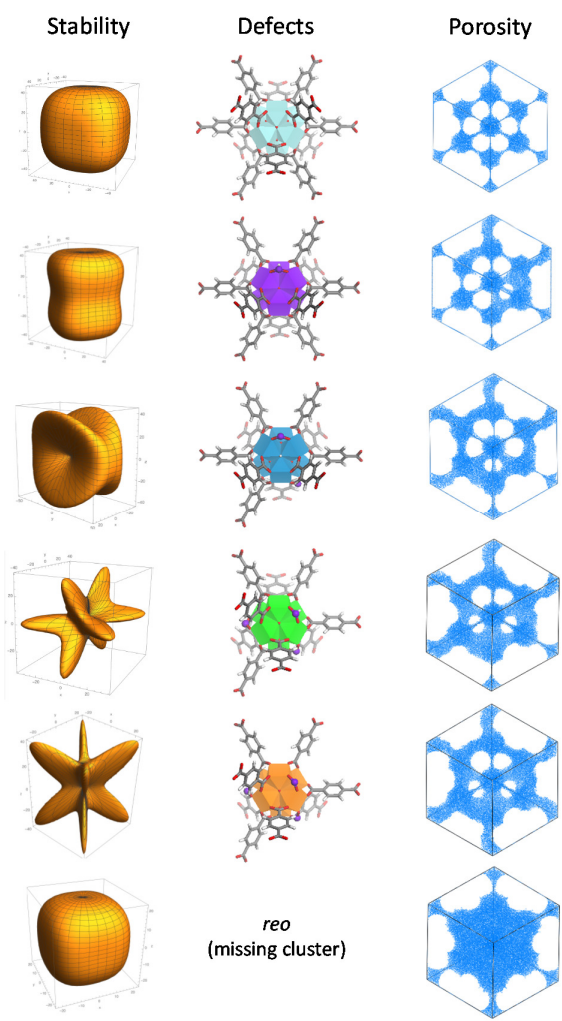

**Figure 1** Snapshot of calculations for stability and porosity on defective structures. Spatially-dependent Young's modulus $E$ is shown to represent a measure of stability. Porosity available for adsorption is shown is blue for a probe diameter of 3 Å. Defects are shown from the perspective of a single cluster where additional linkers are consecutively removed, except for the *reo* structure that is also missing a cluster. Each missing ligand is replaced with either formate, acetate, chloride or hydroxide ions. Each cluster is colored according to its coordination number: 12 – light blue, 11 – purple, 10 – dark blue, 9 – green and 8 – orange.

**Adsorption**

The adsorption of pure $CO_2$, and $N_2$ were simulated using the Grand Canonical Monte Carlo (GCMC) method. The adsorbates were mimicked with a three-site model to account for the quadrupole moment. The C-O bond length in $CO_2$ was 1.18 Å and the bond angle ∠OCO was 180°. The charges on C and O atoms were +0.576e and –0.288e (e = 1.6022×10-19 C the elementary charge), resulting in a quadrupole moment of –1.29×10-39 C·m2. The model reproduced the isosteric heat and isotherm of $CO_2$ adsorption in silicate.[27] $N_2$ had an N-N bond length of 1.10 Å, a charge of -0.482e on the N atom, and a charge of +0.964 at the center-of-mass, which were fitted to the experimental bulk properties of $N_2$.[28] Based on this model, the quadrupole moment of $N_2$ was $-4.67 \times 10^{-40}$ C·m². The potential parameters were adopted from earlier work.[29, 30] The interactions of the gas-adsorbent and gas-gas were modeled as a combination of pairwise site-site Lennard-Jones (LJ) and Coulombic potentials. The LJ potential parameters of the framework atoms were adopted from the Dreiding force field and for the zirconium atom, the Universal force field (UFF)[26] was employed. The cross LJ parameters were evaluated by the Lorentz-Berthelot mixing rules.

The chemical potentials of the adsorbate in the adsorbed and bulk phases are identical at thermodynamic equilibrium, GCMC simulation allows one to relate the chemical potentials of the adsorbate in both phases and has been widely used for the simulation of adsorption. The framework atoms were kept frozen during simulation. This is because adsorption involves low-energy equilibrium configurations and the flexibility of the framework has a marginal effect, particularly on the adsorption of small gases.[31] The LJ interactions were evaluated with a spherical cut-off equal to half of the simulation box with long-range corrections added; the Coulombic interactions were calculated using the Ewald sum method. The number of trial moves in a typical GCMC simulation was $2 \times 10^7$, though additional trial moves were used at high loadings. The first $10^7$ moves were used for equilibration and the subsequent $10^7$ moves for ensemble averages. Four types of trial moves were attempted in GCMC simulation, namely, displacement, rotation, and partial regrowth at a neighbouring position, and entire regrowth at a new position. Unless otherwise mentioned, the uncertainties are smaller than the symbol sizes in the figures presented.

The pore volumes of the adsorbents were obtained according to the thermodynamic method proposed by Myers and Monson.[32] The UFF force field[33] was used to describe the LJ interactions of the framework atoms while the LJ parameters for helium were taken from the work of Talu and Myers.[34] The geometric surface area was calculated from a simple Monte Carlo integration technique where the centre of the mass of the probe molecule with a hard sphere is rolled over the framework surface. In this method, a nitrogen probe molecule was used to calculate the accessible surface area. Program poreblazer_V3.2 was used to calculate the geometric surface and pore volume.[35] The accessible surface area was also calculated based on simulated $N_2$ adsorption isotherms satisfying BET consistency criteria. Supplementary Information Table S1 shows the calculated geometric and accessible surface area, density and pore volume.

**Stability**

Complimentary to the adsorption simulations, the elastic constants were calculated for the defective structures, which were defined at the unit cell scale in the Materials Section, to quantify the mechanical stability and softness. These calculations were based on DFT and performed with the Crystal14 software,[36] adapted to ordered solids as it deals with wave functions built on atom-centered functions (of Gaussian type), and efficiently uses translation invariance as well as point group symmetries. The basis sets used here are identical as those used previously.[37-39] The k-point mesh is generated using the Monkhorst-Pack method.[40] To compute exchange and correlation contribution to the energy, a solid-state adapted exchange-correlation (XC) functional, named PBESOL0, was used – the latter was chosen because of its good performances in modelling cell parameters and mechanical properties of solids in general,[41] and of non-defective UiO-66 in particular.

Within this computation scheme, each structure was first relaxed (both cell parameters and atomic coordinates, while keeping translational invariance and point group symmetries). Once an energy minimum was reached, elastic constants were calculated to quantify the mechanical stability and softness of the defective structures. Calculations were carried out by computing the energy variations subsequent to small deformations of the unit cell, following either of the 6 distinct deformation modes (3 for compression, 3 for shearing) for each mode, deformations corresponding to a strain of ±0.01 were considered (along with the non-defective structure), and geometry optimizations at fixed strain were carried out (keyword used: ELASTCON).

For the elastic constants $C_{ij}$ computed, minimal/maximal values of linear compressibility $\beta$, Young's modulus $E$, shear modulus $G$ and Poisson's ratio $v$ were estimated, using the software Elate[42] (available at http://progs.coudert.name/elate). Spatially-averaged quantities were also calculated based on the Hill's averaging scheme.[43]

## Results and discussion

Simulated $CO_2$ uptake within a range of defective structures except for the *reo* type is shown in Figure 2. There are clear trends observed with the increasing number of missing linkers. At low pressures there is a general decrease in uptake with increased defect concentration. While at high pressures, the opposite trend is observed where an increase in uptake is observed with defect concentration. There is little variation in uptake amongst the structures with different modulators. Formate inhibits the uptake the most at low pressures (1 bar) while chloride helped maintain any loss in uptake. Uptake is sensitive to the interaction energy and surface area at low pressures while at high pressures the uptake is more correlated with pore volume.[44] In this case, structures with formate-capped defects exhibit the lowest surface areas and lowest pore volumes compared with the other types, which explains the lower uptake observed at low pressures (see Table S1).

The effect of *reo* type defective structures is more interesting, see Figure 3. A *reo* type defective structure is considered here with a missing cluster and an average of 8 linkers per cluster (note that the average number of missing linkers is dependent on the number of missing clusters per unit cell). For comparison, a structure also with 8 linkers per cluster but without a missing cluster is considered. The *reo* type defective structure consists of a large cavity of diameter 17 Å compared to the cavity diameter of 8.9 Å formed by only missing linkers, as shown in Figure 1 and pore size distributions in Fig. S6 of Supporting Information. Interestingly the uptake is dramatically increased by more than double at large pressures for the *reo* scenario. This is impressive, but the question arises: Do these enhancements come with a compromise in stability?

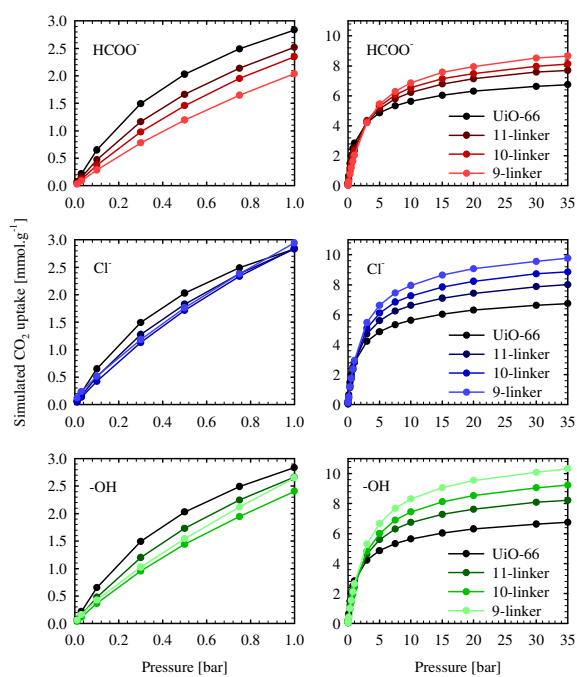

**Figure 2** Simulated $CO_2$ uptake at low (left) and high (right) pressures within a range of defective scenarios.

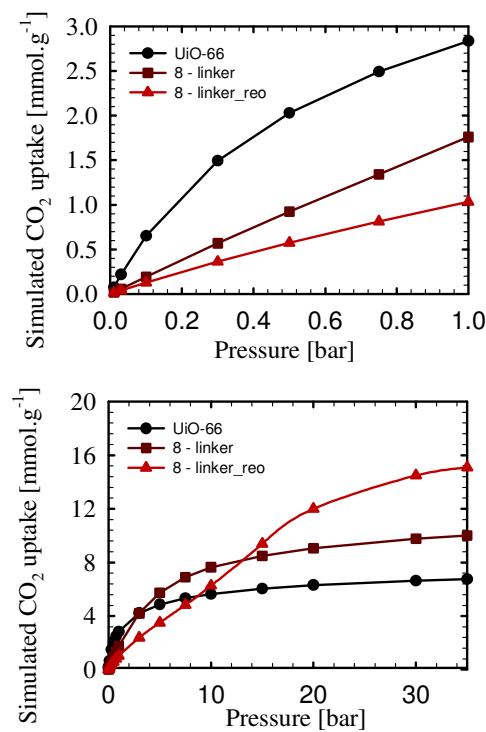

**Figure 3** Simulated $CO_2$ uptake at high pressures within a *reo* type defective structure, a structure with 8 linkers per cluster on average and a perfect *fcu* structure.

Note that the *reo* scenario investigated above considers the case of one missing cluster which represents a homogeneous case with equally spaced defects. In reality, the defects may be disordered or correlated. Here six scenarios are considered, see Figure 4 with clusters color coded according to their coordination number. All six scenarios have an average of 9 linkers per cluster. For the first, every cluster has exactly 9 linkers. The second scenario has a combination of clusters with 8 and 10 linkers. The remaining scenarios are based on *reo* type defects where each cell has 4 missing clusters with varying degree of adjacency from 1 to 4. Adjacency is defined here as the number of missing clusters directly adjacent from each other. For example, an adjacency of 2 means that there are 2 missing cluster sites directly next to each other. The construction of these scenarios is an attempt to represent vacancy correlation, similar to that observed experimentally by Cliffe e*t al.*[17]

Simulated $CO_2$ uptake is compared within all scenarios. For clarity, Figure 4 depicts isotherms for only three scenarios. The isotherms within all *reo* type scenarios were almost identical. The dominant (most common) cavity size and pore volume remained unchanged which is why there no difference in the isotherms. Once again there is higher uptake observed within *reo* type defective structures compared to the missing ligand scenarios. There is a slight increase in uptake with less symmetry, i.e. the case of mixed clusters with z = 8 and 10. Pore size distributions in Fig. S6 of Supporting Information, show that the pore size shifts only from 8.6 to 8.9 Å from the perfect *fcu* structure.

Stability is analysed by considering the elastic constants generated from DFT-based simulations, as described in the Experimental Section. It can be seen by the maximum and minimum elastic moduli in Table 1 for formate-modulated defective structures that, upon increasing the proportion of defects, anisotropy in mechanical properties is increased significantly. This anisotropy is amplified graphically in Figure 1 and a closer view in Fig. S7 of Supporting Information for the Young's modulus. Covering the Young's and shear modulus, this increase is (in orders of magnitude) exponential with the defect number per cell $n = 24 - 2z$, with the corresponding anisotropy factors like $E_{max}/E_{min}$ reaching values > 50 for $z = 8$. An exception is the *reo* structure, which is of much higher symmetry than the other defective structures.

**Table 1** Minimal/maximal values of linear compressibility $\beta$, Young's modulus $E$ and shear modulus $G$ (GPa). Defective structures listed here are based on the formate modulator.

| Structure | $\beta_{min}$ | $\beta_{max}$ | $E_{min}$ | $E_{max}$ | $G_{min}$ | $G_{max}$ |
| --- | --- | --- | --- | --- | --- | --- |
| *fcu* | 7.91 | 7.91 | 45.18 | 54.62 | 17.10 | 21.27 |
| z = 11 | 4.66 | 11.95 | 33.12 | 51.31 | 12.94 | 21.29 |
| z = 10 | -29.44 | 109.3 | 4.88 | 50.95 | 2.48 | 21.41 |
| z = 9 | -30.99 | 94.51 | 2.27 | 50.4 | 0.76 | 21.37 |
| z = 8 | -31.7 | 91.73 | 0.94 | 50.23 | 0.25 | 21.81 |
| *reo* | 16.94 | 16.94 | 22.52 | 27.28 | 8.60 | 10.75 |

Minimal Young's and shear moduli are obtained for axes playing a specific role in the structure, e.g. for $z = 10$ the easiest compression is in the direction parallel to missing linkers (*x*-axis, see Fig. S4 in Supporting Information) and the easiest shearing is that of the ($[10\bar{1}]^\perp$) plane along the [101] direction. Note that one observes negative linear compressibility (directions for which $\beta < 0$, i.e. along which the system contracts under uniform compression) and negative Poisson' ratio, for $z \leq 10$ (see SI).

This auxetic behaviour (compression along a given direction induces cell shrinking along a transverse direction) has been observed computationally in a number of MOFs including the 'wine-rack' series of MIL-140A, MIL-122(In), MIL-53(Al), MIL-53(Ga) and MIL-47,[42] along with square-shaped DMOF-1[45] and the orthorhombic ZIF-4.[46] This property is thought to give rise to indentation resistance and fracture toughness.[47] This effect could also explain the defect-dependent increase in negative thermal expansion observed experimentally for hafnium-based UiO-66.[48]

Modulator-dependent mechanical properties are considered here for structures with the same topology ($z = 8$ and $z = 10$) but with different substituents, namely formate ($HCOO^-$), acetate ($H_3CCOO^-$), hydroxyl ($H_2O, HO^-$), chloride ($H_2O, Cl^-$) and trifluoroacetate ($F_3CCOO^-$) listed in Table 2. While the bulk modulus is almost insensitive to the nature of the substituent, concerning Young's and shear moduli one finds much more contrast between structures, which can be grouped in two sets.

On one hand, formate, hydroxyl and chloride substituents give values close to each other: $E \cong 18$ GPa, $E_{min} \cong 0.8$ GPa, $G \cong 7$ GPa and $G_{min} \cong 0.2$ GPa. Note the very small $G_{min}$ value while $E_{min}$ is about 4 times higher. Take for instance the formate-substituted structure (see Fig. S4-right): here $(xy)=[001]^\perp$ is the plane common to all missing BDC's outgoing from a given cluster, so that each cluster has 4 outgoing BDC's in the $[110]^\perp$ plane and 4 in the $[1\bar{1}0]^\perp$ plane. In this context, shearing can be exerted on e.g. $[110]^\perp$ without stretching the first group of ligands, and with minimal stretching ($\propto \epsilon^2$ with $\epsilon$ the shearing strain) on the other ones. The very same mechanism makes compression easiest along the *x* and *y* axis (or their counterparts in the various structures, see Axis 3 in Table 2). For comparison, in the non-depleted *fcu* structure, the 4 additional BDC's (absent here) would have been stretched proportionally to $\epsilon$ (2 elongated and 2 compressed), explaining the much higher (by 2 orders of magnitude) values of $G_{min}$ and $E_{min}$ – see for *fcu* in Table 1.

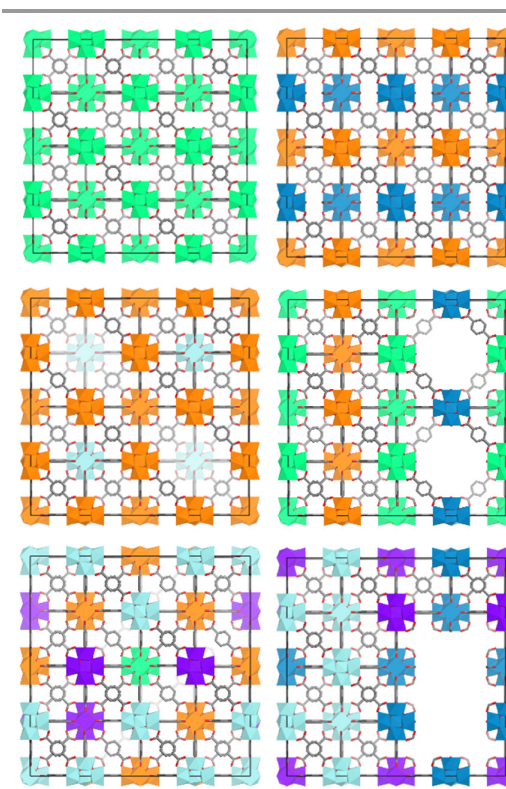

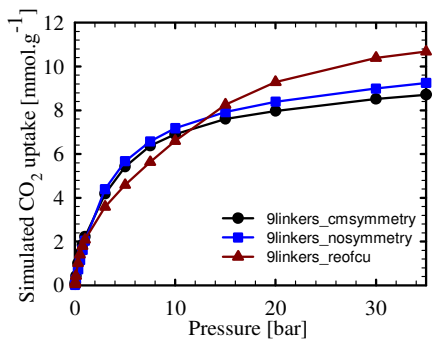

**Figure 4** (Top) Defect scenarios where all structures have the same average number of linkers per cluster ($z = 9$) and varying heterogeneity. Each cluster is colored according to its coordination number: 12 – light blue, 11 – purple, 10 – dark blue, 9 – green and 8 – orange. (Bottom) Simulated $CO_2$ uptake in three generalised scenarios: with symmetry, with lower symmetry and a combination of perfect *fcu* and defective *reo*.

The other set contains acetate and trifluoroacetate-substituted structures. Here, Young's and shear moduli are much larger than for other substituents, by an order of magnitude. While $G_{min}$ still corresponds to the same kind of shearing, it is rendered less efficient (more costly) by the larger size taken by acetate and trifluoroacetate ions. Indeed, in the optimized structure methyl groups of both acetate ions replacing a missing BDC are relatively close to each other (H–H and C–C distances of about 2.5 Å and 3.3 Å respectively); this makes the repulsion between their corresponding electronic clouds non-negligible, such that these ions can't be brought closer to each other as easily as in the case of e.g. formate substituents. Eventually the negative linear compressibility is observed for all 4 types of substituents (along the direction normal to the plane of missing bdc's), the value $-\beta_{min} = |\beta_{min}|$ (not shown) is, however, about 3 times larger for small substituents than for acetate and trifluoroacetate, for the same reasons as discussed above.

**Table 2** Bulk modulus $B$, and spatially-averaged Young's modulus $E$ and shear modulus $G$ with minimal values (GPa). Four versions of $z = 8$ structures with different types of substitution ligands (or modulators) are listed. Values for formate ($z = 10$) and *fcu* structures differ slightly from those in Table 1, because of different convergence criteria in $C_{ij}$ calculations.

| Modulator | $B$ | $E$ | $E_{min}$ | $G$ | $G_{min}$ |
|---|---|---|---|---|---|
| *8 linkers per cluster* | | | | | |
| Formate | 19.04 | 18.88 | 0.78 | 7.07 | 0.21 |
| Acetate | 17.33 | 25.18 | 7.56 | 10.01 | 2.73 |
| Hydroxyl | 17.64 | 18.24 | 0.85 | 6.87 | 0.23 |
| Chloride | 17.82 | 18.15 | 0.77 | 6.82 | 0.2 |
| *10 linkers per cluster* | | | | | |
| Formate | 22.82 | 30.53 | 4.85 | 11.95 | 2.47 |
| Acetate | 30.25 | 37.25 | 13.71 | 14.39 | 6.05 |
| Trifluoroacetate | 33.16 | 38.22 | 19.45 | 14.61 | 7.79 |
| *12 linkers per cluster* | | | | | |
| *fcu* | 42.13 | 50.61 | 44.98 | 19.47 | 17.01 |

In comparison, Figure 5 displays the spatially-averaged constants along with simulated $CO_2$ uptake. The spatially-averaged mechanical constants are less dramatic in variation with the defect concentration compared with the change in minimum and maximum values from Table 1. The trend is rather expected, that the structures get softer (smaller $B$, $E$ and $G$) when the number/concentration of defects increases. Furthermore, Young's modulus increases along with uptake at low pressure which could have important benefits for $CO_2$ capture separations.[49] For high pressures, the spatially-averaged Young's modulus decreases with increasing uptake, meaning that there is a compromise between uptake and stability. This is understandable given the general trend that missing linkers results in less structural support comprising mechanical stability but also resulting in more pore volume that can enhance $CO_2$ uptake. Remarkably, *reo* type defective structures fight against this trend and offer a more stable structure coupled with an increase in uptake at high pressures.

While this study has covered a wide scope of defect scenarios there are many questions that still remain for future work. For example, the stability calculations are computationally expensive and incapable of exploring heterogeneity on a large scale such as the supercells described in Figure 4. One can not take a simple average of elastic constants over a large scale. Heterogeneity is in itself a wide area of study which is covered here in a limited fashion. Effects of crystal edges are also not taken into account that have significant consequences on adsorption.[50] Furthermore, stability is treated here as independent of adsorption, which is likely to play a role. Nonetheless, this study merges multiple calculations of multiple scenarios in an attempt to bridge the gap of understanding on the compromise between adsorption and stability with defect engineering.

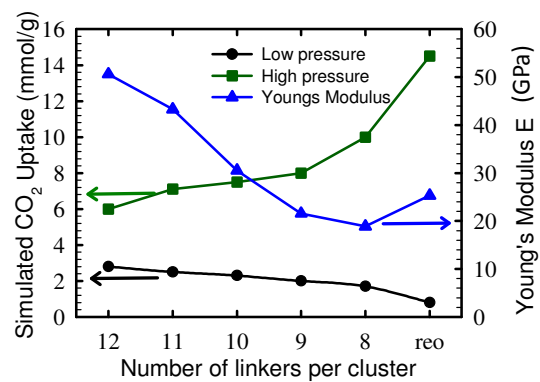

**Figure 5** Simulated $CO_2$ uptake at low (1 bar) and high (30 bar) pressures with Young's modulus as a function of defect types.

## Conclusions

The role of defects on $CO_2$ adsorption capacity and mechanical stability was computationally explored. A systematic variation of defect types were considered including modulator-dependence (formate, acetate, hydroxyl, chloride and trifluoroacetate), defect concentration (8 to 12 linkers per cluster and *reo* type defective structures with missing clusters) and heterogeneity (adjacency of missing clusters).

$CO_2$ uptake at low pressures decreased as a function of defect concentration while uptake increased at high pressures. Uptake did not vary much with varying modulator, although chloride helped maintain any loss in uptake at low pressures.

Heterogeneity was explored by considering the level of adjacency among *reo* type defective structures. The correlation of defects did not effect the $CO_2$ uptake. However, the large cavities and pore volumes led to higher uptake than the perfect *fcu* structure.

The mechanical calculations revealed an especially interesting increase in anisotropy with defect concentration. In contrast, *reo* type defective structures were highly symmetric. Interestingly, acetate and even more so trifluoroacetate substitution was shown to strengthen the UiO-66 an order of magnitude above the other modulators.

Auxeticity (where a system contracts, under directional compression, in at least one transverse direction) was also observed for coordination numbers $z \leq 10$. This effect could lead to superior indentation resistance and fracture toughness.

Finally, stability is compromised at high pressures where uptake is enhanced with an increase in defect concentration.

However, *reo* type defective structures and structures capped with trifluoroacetate maintain relative stability above other type of defective structures.

In summary, stability is compromised when enhancing adsorption through defects but some stability can be maintained by engineering the type and distribution of defects. These results can be considered when designing defective structures for a wide scope of other adsorption applications such as hydrogen and methane storage.

## Acknowledgements

The authors thank Matt Cliffe, Andrew Goodwin, Deanna D'Alessandro and Weibin Liang for discussions and comments. AWT acknowledges the CSIRO Office of the Chief Executive's Julius Award for support. This work was performed using high-performance computing resources from GENCI (Grant x201587069) and with assistance from the National Computational Infrastructure (NCI), which is supported by the Australian Government. This research is also supported by the Science and Industry Endowment Fund.

## Notes and references